\documentclass[%
 reprint,
 superscriptaddress,
 bibnotes,
 amsmath,amssymb,
 aps,
 prb,
 floatfix,
]{revtex4-2}

\usepackage{graphicx}
\usepackage{dcolumn}
\usepackage{bm}
\usepackage{hyperref}
\usepackage{float}
\hypersetup{
  colorlinks   = true,
  urlcolor     = black,
  linkcolor    = black,
  citecolor   = blue
}
\usepackage{etoolbox}
\apptocmd{\sloppy}{\hbadness 10000\relax}{}{}

\usepackage{xcolor}

\newcommand{\ttg}{\ensuremath{t_{2g}}}
\newcommand{\eg}{\ensuremath{e_{g}}}

\newcommand{\JH}{\ensuremath{J_{\text{H}}}}
\newcommand{\tus}[1]{\textsuperscript{#1}}
\newcommand{\tds}[1]{\textsubscript{#1}}
\newcommand{\KRC}{K\tds{2}RuCl\tds{6}}
\newcommand{\CRO}{Ca\tds{2}RuO\tds{4}}

\begin{document}

\title{Nonmagnetic \texorpdfstring{$J = 0$}{J=0} State and Spin-Orbit Excitations in \texorpdfstring{\KRC}{K2RuCl6}}

\author{H. Takahashi}
\affiliation{Max-Planck-Institut f\"{u}r Festk\"{o}rperforschung, Heisenbergstra\ss e 1, D-70569 Stuttgart, Germany}
\affiliation{Department of Physics, The University of Tokyo, Bunkyo-ku, Tokyo 133-0022, Japan}
\author{H. Suzuki}
\email[]{Corresponding author. hakuto.suzuki@tohoku.ac.jp}
\affiliation{Max-Planck-Institut f\"{u}r Festk\"{o}rperforschung, Heisenbergstra\ss e 1, D-70569 Stuttgart, Germany}
\author{J. Bertinshaw}
\affiliation{Max-Planck-Institut f\"{u}r Festk\"{o}rperforschung, Heisenbergstra\ss e 1, D-70569 Stuttgart, Germany}
\author{S. Bette}
\affiliation{Max-Planck-Institut f\"{u}r Festk\"{o}rperforschung, Heisenbergstra\ss e 1, D-70569 Stuttgart, Germany}
\author{C. M{\"u}hle}
\affiliation{Max-Planck-Institut f\"{u}r Festk\"{o}rperforschung, Heisenbergstra\ss e 1, D-70569 Stuttgart, Germany}
\author{J. Nuss}
\affiliation{Max-Planck-Institut f\"{u}r Festk\"{o}rperforschung, Heisenbergstra\ss e 1, D-70569 Stuttgart, Germany}
\author{R. Dinnebier}
\affiliation{Max-Planck-Institut f\"{u}r Festk\"{o}rperforschung, Heisenbergstra\ss e 1, D-70569 Stuttgart, Germany}
\author{A. Yaresko}
\affiliation{Max-Planck-Institut f\"{u}r Festk\"{o}rperforschung, Heisenbergstra\ss e 1, D-70569 Stuttgart, Germany}
\author{G. Khaliullin}
\affiliation{Max-Planck-Institut f\"{u}r Festk\"{o}rperforschung, Heisenbergstra\ss e 1, D-70569 Stuttgart, Germany}
\author{H. Gretarsson}
\affiliation{Deutsches Elektronen-Synchrotron DESY, Notkestrstra\ss e 85, D-22607 Hamburg, Germany}
\author{T. Takayama}
\affiliation{Max-Planck-Institut f\"{u}r Festk\"{o}rperforschung, Heisenbergstra\ss e 1, D-70569 Stuttgart, Germany}
\affiliation{Institute for Functional Matter and Quantum Technologies, University of Stuttgart, Pfaffenwaldring 57, 70569 Stuttgart, Germany}
\author{H. Takagi}
\affiliation{Max-Planck-Institut f\"{u}r Festk\"{o}rperforschung, Heisenbergstra\ss e 1, D-70569 Stuttgart, Germany}
\affiliation{Department of Physics, The University of Tokyo, Bunkyo-ku, Tokyo 133-0022, Japan}
\affiliation{Institute for Functional Matter and Quantum Technologies, University of Stuttgart, Pfaffenwaldring 57, 70569 Stuttgart, Germany}
\author{B. Keimer}
\email[]{Corresponding author. B.Keimer@fkf.mpg.de}
\affiliation{Max-Planck-Institut f\"{u}r Festk\"{o}rperforschung, Heisenbergstra\ss e 1, D-70569 Stuttgart, Germany}

\date{\today}

\begin{abstract}
Spin-orbit Mott insulators composed of $\ttg^4$ transition metal ions may host excitonic magnetism due to the condensation of spin-orbital $J=1$ triplons. Prior experiments suggest that the $4d$ antiferromagnet \CRO\ embodies this notion, but a $J = 0$ nonmagnetic state as a basis of the excitonic picture remains to be confirmed. We use Ru $L_3$-edge resonant inelastic x-ray scattering to reveal archetypal $J$ multiplets with a $J=0$ ground state in the cubic compound \KRC, which are well described within the $LS$-coupling scheme. This result highlights the critical role of unquenched orbital moments in $4d$-electron compounds and calls for investigations of quantum criticality and excitonic magnetism on various crystal lattices.
\end{abstract}

\maketitle

In the presence of strong spin-orbit coupling (SOC), correlated electron systems have been predicted to display a rich set of emergent phenomena \cite{Witczak-Krempa:arcmp:2014,Rau:arcmp:2016,Takayama:jpsj:2021} ranging from spin liquid behavior \cite{Winter:jpcm:2017,Hermanns:arcmp:2018,Takagi:natrp:2019} to Weyl semimetals \cite{Hasan:arcmp:2017,Yan:arcmp:2017,Nagaosa:natrm:2020}. A prime platform for the experimental exploration of these predictions is 4$d$ transition metal compounds with the orbital degrees of freedom, where the strength of the intra-ionic SOC is comparable to the inter-ionic exchange interactions.
At the same time, experimental work on these materials has brought about a string of remarkable discoveries that had not been anticipated by theory. A prominent example is an enigmatic nonmagnetic insulating state that was discovered in a variety of different compounds based on Ru$^{4+}$ ions (electron configuration 4$d^4$), which are expected to carry $S=1$ spins according to Hund’s rule. Models proposed to account for the nonmagnetism in these systems generally invoke quantum singlet formation in the spin degrees of freedom, such as exchange-coupled spin-1 chains with collective-singlet ground states \cite{Lee:natm:2006} or spin-singlet dimers generated by orbital ordering \cite{Miura:jpsj:2007,Yun:prb:2019,Khalifah:sci:2002,Wu:prl:2006}. However, singlet ground states can also be generated by the antiparallel alignment of the spin and orbital angular momenta promoted by the intra-atomic SOC \cite{Khaliullin:prl:2013}. In principle, these different mechanisms leave highly specific fingerprints on low-energy excitations such as magnons and spin-orbit excitons. Neutron scattering experiments on the antiferromagnetic Ru$^{4+}$ compound \CRO\ have indeed revealed a soft longitudinal magnon mode that has been interpreted as a hallmark of a proximate quantum phase transition into a SOC-driven singlet state \cite{Jain:natp:2017}. In compounds with nonmagnetic ground states \cite{Lee:natm:2006,Miura:jpsj:2007,Yun:prb:2019,Khalifah:sci:2002,Wu:prl:2006}, however, materials issues (including especially the lack of sizeable, disorder-free single crystals for neutron experiments) have precluded spectroscopic measurements that could definitively resolve the origin of quantum singlet formation.

Here we report Ru $L_3$-edge resonant inelastic x-ray scattering (RIXS) experiments on single crystals of \KRC\ \cite{Johannesen:ic:1963}, a compound where Ru$^{4+}$ ions are arranged in a cubic crystal field environment without any distortions. The spin-orbit multiplet excitations revealed by RIXS yield a definitive spectroscopic fingerprint of a SOC-driven $J=0$ singlet state. Our results point to spin-orbit entanglement as a generic characteristic of insulating Ru$^{4+}$ compounds with widely different lattice architectures. In the light of prior theoretical work, \KRC\ can be regarded as the parent compound of a class of “excitonic magnets” \cite{Khaliullin:prl:2013,Anisimov:prl:2019,Chaloupka:prb:2019}, in which the inter-ionic exchange interaction drives the condensation of $J=1$ spin-orbit excitons.
The theoretically predicted quantum phase transition separating these two regimes will be an interesting subject of further research.

\begin{figure}[ht]
  \centering
  \includegraphics[angle = 0, width = 0.48\textwidth, clip=true]{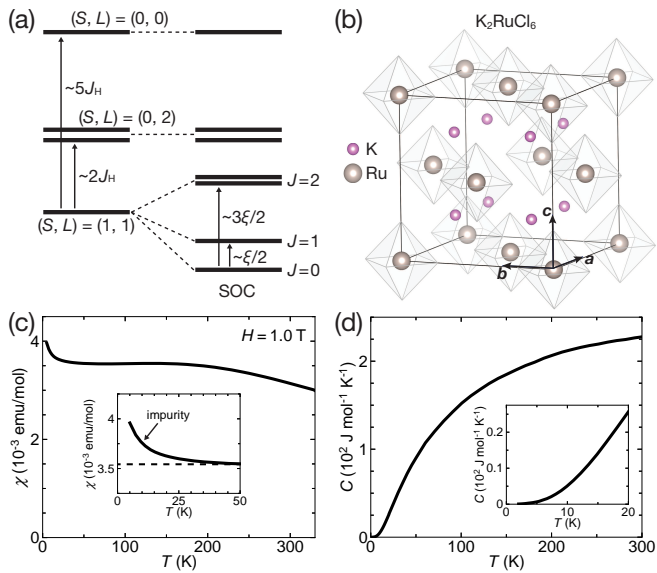}
  \caption{\label{fig:1} (a) Energy levels of $\ttg^4$ electron configurations with Coulomb interactions and spin-orbit coupling (SOC).
  The left column shows the $(S,L)$ multiplets formed by the Coulomb interactions. The lowest $(S,L)=(1,1)$ multiplets are further split by the SOC into $J$ multiplets (right column).
  (b) Crystal structure of \KRC. Ru\tus{4+} ions are octahedrally coordinated by Cl\tus{-} ions (grey octahedra), and the RuCl\tds{6} units form the face-centered-cubic lattice.
  (c) Temperature dependence of the magnetic susceptibility $\chi(T)$ of \KRC\ powder measured with an applied magnetic field of 1.0 T, showing a plateau characteristic of the Van Vleck susceptibility. The inset shows an expanded plot below 50 K. The upturn is attributed to impurities.
  (d) Temperature dependence of the specific heat $C(T)$. The inset shows an expanded plot below 20 K.
  }
\end{figure}

Before presenting the RIXS data, we briefly review the energy levels of a $4d^4$ electron system in a cubic crystal field, which generates a gap between the \ttg\ and \eg\ orbitals. Figure \ref{fig:1}(a) presents the energy levels of the $\ttg^4$ electron configuration in the $LS$-coupling scheme, where the Hund's coupling \JH\ outweighs the SOC constant $\xi$. Without SOC, the states are labeled by the total spin $S$ and effective orbital angular momentum $L$. The ground states are given by the $(S, L)=(1, 1)$ states, and the $(0, 2)$ and $(0, 0)$ excited states are located at the energy of $\sim 2\JH$ and $\sim 5\JH$, respectively. The orbital degeneracy of the (1,1) states is commonly lifted via the Jahn-Teller (JT) distortion of the octahedra \cite{Bersuker::2006}. However, in the presence of SOC expressed as $(\xi/2) \bm{S} \cdot \bm{L}$, the $(S, L)=(1,1)$ states are split into spin-orbital $J$ multiplets. The ground state is given by the nonmagnetic $J = 0$ singlet with suppressed JT effect, and the $J = 1,2$ excited states are located at the energy of $\sim\xi/2$ and $\sim3\xi/2$, respectively. The hierarchy of these excited states can thus serve as a diagnostic fingerprint of the ground state.

Figure \ref{fig:1}(b) shows that \KRC\ has a double-perovskite-like crystal structure, where the RuCl\tds{6} octahedra form a face-centered-cubic (fcc) lattice. Our powder x-ray diffraction measurements reveal that it does not show any spontaneous lattice distortions that break the cubic symmetry down to low temperature \cite{SM}. Since the distance between the Ru ions is as long as 6.8 \AA, Ru $d$ orbitals overlap only weakly (as evidenced by band structure calculations \cite{SM}), and magnetic interactions can be assumed to be small. Indeed, the magnetic susceptibility of \KRC\ shown in Fig. \ref{fig:1}(c) manifests no anomaly down to 5 K, excluding the presence of magnetic ordering. Instead, it exhibits a plateau below $\sim$ 200 K, characteristic of the Van Vleck paramagnetic susceptibility of isolated $J=0$ ions. A subtle upturn at low temperatures (see inset) can be attributed to the Curie paramagnetic susceptibility of impurities;  the Curie fit yields an impurity concentration of only $\sim$ 0.2 \%, demonstrating the high purity of our \KRC\ samples. The specific heat data [Fig. \ref{fig:1}(d)] shows no anomaly down to 2 K and exhibits an insulating behavior with standard phonon contribution $\propto T^3$ at low temperatures. Also, the optical conductivity indicates that \KRC\ is a large-gap insulator without impurity-induced in-gap states \cite{Rabinovich:unp:}.
These properties make \KRC\ a more ideal platform to investigate the $J$ multiplets than the 5$d^4$ iridium double perovskites, where the magnetic correlations of impurities have aroused controversy over the nature of their magnetism (see \cite{Fuchs:prl:2018} and references therein).

\begin{figure}[H]
  \centering
  \includegraphics[angle = 0, width = 0.48\textwidth, clip=true]{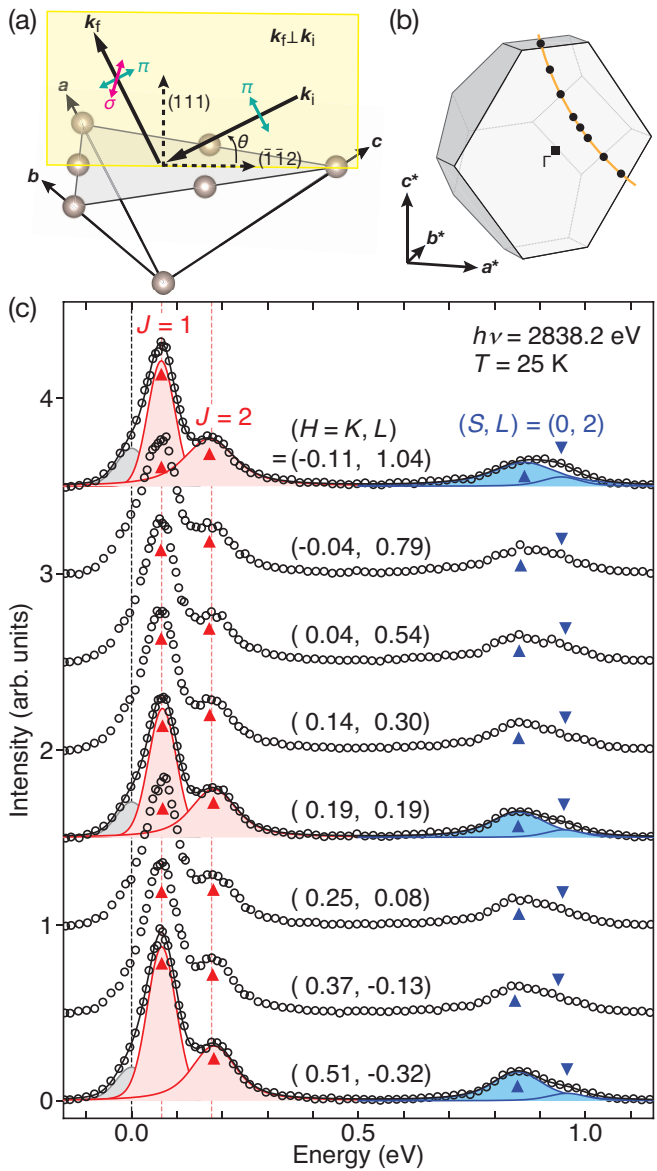}
  \caption{\label{fig:2} (a)  Scattering geometry of the RIXS experiment. The \KRC\ crystal has a natural (111) surface (grey) and the scattering plane (yellow) contains the (111) and ($\bar{1}$$\bar{1}$2) directions.
  The scattering angle is fixed at 90$^{\circ}$. The incident x-ray photons with momenta ${\bm k}_i$ are linearly $\pi$-polarized and the polarizations of the scattered photons with momenta ${\bm k}_f$ are not analyzed. The momentum transfer ${\bm q}={\bm k}_i-{\bm k}_f$ is scanned by changing the sample angle $\theta$. (b) The measurement path of the RIXS experiment in the fcc Brillouin zone. The black points correspond to the measured momenta. (c) Ru $L_3$ edge (2838.2 eV) RIXS spectra of \KRC\ taken at 25 K. Overlaid are the examples of spectral decompositions into five Voigt functions, representing elastic scattering (grey), excitations to the $J$ multiplets (red), and excitations to the $(S,L) = (0,2)$ states (blue). Red and blue triangles indicate the determined peak positions. The dotted red lines show the statistical average of peak positions of the $J$ multiplets (66 and 176 meV).\\
  \,
  }
\end{figure}

\textit{RIXS experiment.}--- Measurements were carried out using the intermediate-energy RIXS (IRIXS) spectrometer at the Dynamics Beamline P01 of PETRA III, DESY \cite{Gretarsson:jsr:2020}. The scattering geometry of the RIXS experiment is illustrated in Fig. \ref{fig:2}(a). The incident x-ray energy was tuned to the Ru $L_3$ absorption edge ($\sim$2838 eV). The \KRC\ single crystal was mounted so that the scattering plane (yellow) contains the (111) and ($\bar{1}$$\bar{1}$2) directions \cite{SM}. The incident x-ray photons with momenta ${\bm k}_i$ were $\pi$-polarized. Scattered photons with momenta ${\bm k}_f$ were detected at a fixed scattering angle of 90$^\circ$ using a $\mathrm{SiO_2}$ ($10\bar{2}$) diced spherical analyzer and a CCD camera. The polarizations of the outgoing photons were not analyzed. The incident x-ray beam was focused to a beam spot of 20$\times$150 $\mu$m$^{2}$ (H $\times$ V). The momentum transfer ${\bm q} = {\bm k}_i - {\bm k}_f$ was varied by changing the sample angle $\theta$ from 22${}^\circ$ to 72${}^\circ$, and the measured momenta in the fcc Brillouin zone are indicated in Fig. \ref{fig:2}(b). In the following, the momentum transfer is expressed in reciprocal lattice units (r.l.u.). The exact position of zero energy loss and the energy resolution of $\Delta E \sim 90$\ meV were determined by the center and the full-width at half-maximum, respectively, of the nonresonant elastic signal from silver paint deposited next to the sample.

\textit{Results.}---
Figure \ref{fig:2}(c) shows Ru $L_3$ edge (2838.2 eV) RIXS spectra of \KRC\ taken at 25 K with different momentum transfer ${\bm q}$. The incident x-ray energy was selected to maximize the intensity of spectral features below 1 eV.
The correction for x-ray self-absorption has been applied to the spectra following the procedure in Ref. \cite{Minola:prl:2015}.
For all the measured momenta, one observes two sharp low-energy peaks below 0.3 eV and a broad asymmetric peak around 0.9 eV. These excitations do not show energy dispersions within the experimental precision, confirming negligible intersite exchange interactions between the $J$ multiplets. The absence of energy dispersion is contrasted with the propagation of the $J$ excitons in 5$d$ electron Ir double perovskites \cite{Kusch:prb:2018}. Furthermore, the weak intensity modulation of the $J$ multiplets as a function of ${\bm q}$ is in striking contrast to the strong intensity modulation of the $J$ multiplets in Ca$_2$RuO$_4$ \cite{Gretarsson:prb:2019}, indicating highly symmetric spin-orbital wavefunctions in the cubic \KRC.

To quantitatively extract the energies of individual excitations, we performed spectral fitting of the RIXS data. First, the low-energy region (-0.15, 0.4) eV was fitted by three Voigt profiles: elastic scattering fixed at 0 eV (grey) and the $J = 1, 2$ transitions (red) at $\sim 70$ meV and $\sim 180$ meV. The energy ratio of these two transitions is 2.6, which is close to 3 expected for the $J = 1, 2$ multiplets in the $LS$-coupling limit.
Next, the broad peak around $0.9$ eV is decomposed into two Voigt functions (blue) centered at $\sim$ 850 meV and $\sim$ 950 meV.
These peaks are readily assigned to the excitations from $J = 0$ to the $(S,L) = (0,2)$ states. Note here that these states are in fact composed of a $T_{2g}$ triplet and an $E_g$ doublet

{\setlength{\parindent}{0cm}
under a cubic crystal field. Likewise, the $J = 2$ "quintet" is composed of a $T_{2g}$ triplet and an $E_g$ doublet under the cubic crystal field, but as their splitting is of the second order in the SOC \cite{Takayama:jpsj:2021}, it is not resolved in the present measurement.
Since the observed multiplets do not exhibit energy dispersion within the experimental precision, we take statistical averages of the peak positions from different ${\bm{q}}$ data as the local excitation energies. In the following, we use $66\pm 2$ meV ($J=1$), $176\pm 4$ meV ($J=2$), $854\pm 7$ meV and $948\pm 7$ meV [$(S,L) = (0,2)$] for each observed excitation, where the errors represent the standard deviation of the peak positions.
}

\textit{Discussion.}---
Based on the determined energies of the intraionic excitations, we estimate microscopic multiplet parameters of the $4d^{4}$ ions in K$_2$RuCl$_6$.
The ionic model Hamiltonian for the $4d$ electrons in a Ru\tus{4+} ion under the cubic crystal field consists of the intra-atomic Coulomb interaction terms in the Kanamori form, $H_\mathrm{C}$ \cite{Sugano::1970,Georges:arcmp:2013}, the intra-atomic SOC, $H_\mathrm{SOC}$, and the cubic crystal field $H_\mathrm{cub}$ \cite{SM}.
We employ the spherical symmetry approximation of the interaction terms in $H_\mathrm{C}$, which are then parameterized with Racah parameters \cite{Sugano::1970}. This approximation imposes the condition $U'_{mm'} = U - 2J_{\mathrm{H},mm'}$, and the result is independent of $U$ in the ionic model with a fixed electron number.
Free parameters to be determined are thus the Hund's coupling between the \ttg\ orbitals \JH, the SOC $\xi$, and the crystal field strength $10Dq$. The RIXS transition amplitude from the ground state was calculated within the dipole approximation and fast collision approximation \cite{Luo:prl:1993,Veenendaal:prl:2006}.

\begin{figure}[tb]
  \centering
  \includegraphics[angle = 0, width = 0.48\textwidth, clip=true]{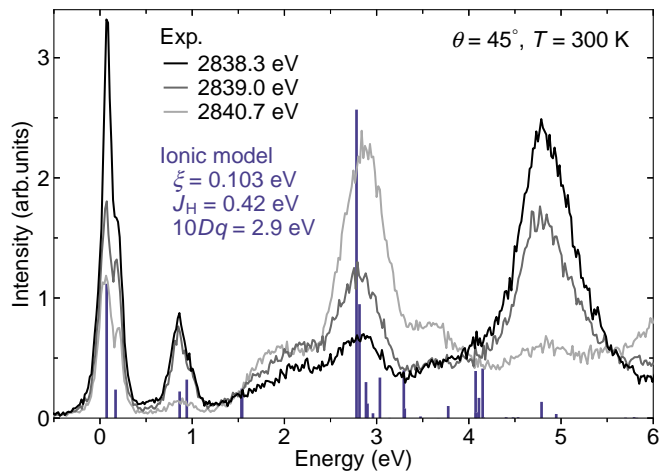}
  \caption{\label{fig:3} High-energy RIXS spectra of \KRC\ taken with different incident x-ray photons of 2838.3 eV (black), 2839.0 eV (dark grey), and 2840.7 eV (light grey). The measurements were performed at $\theta = 45^\circ$ at 300 K.
  The blue bars show the theoretical RIXS intensity for the ionic model Hamiltonian \cite{SM} with the optimal parameters $\xi = 0.103$, $\JH = 0.42$, and $10Dq = 2.9$ eV.
  }
\end{figure}

Figure \ref{fig:3} compares the theoretical RIXS spectra for the optimal parameters $\xi = 0.103$, $\JH = 0.42$, and $10Dq = 2.9$ eV \cite{SM} (blue bars) and the experimental RIXS spectra at 300 K measured with different incident energies [2838.3 (black), 2839.0 (dark grey), and 2840.7 (light grey) eV] in a wide energy region.
Upon increasing the incident x-ray energy, the spectral features below 1 eV are diminished, while the peak around 2.9 eV is strongly enhanced. This suggests that the former originate from excitations within the $\ttg^4$ electron configuration and the latter from transitions to the $\ttg^3\eg$ configuration. With the optimal parameter set, the calculation captures the energies of the $J$ multiplets, the $(S, L) = (0, 2)$ states, and the $\ttg\rightarrow\eg$  crystal field transitions. It also reproduces the intensity ratio between the $J = 1$ and $J = 2$ transitions, where the $J=1$ dipole transition is enhanced in the 90$^{\circ}$ scattering geometry. The weak transition at 1.5 eV is the excitation to the high-spin $S = 2$ multiplet with the $\ttg^3 \eg$ configuration. We note that the transition to the $(S, L) = (0, 0)$ state at 1.8 eV is prohibited by the selection rule \cite{Ament:rmp:2011}.
The peak around 5 eV is assigned to the charge-transfer excitations from the Cl $p$ orbitals to the Ru \ttg\ orbitals. The features above $\sim$ 1.5 eV bear resemblance to those observed in $\alpha$-RuCl$_3$ \cite{Suzuki:arxiv:2020}, whereas the low-energy multiplets are quite different reflecting the distinct $d^4$ ($J=0$) vs $d^5$ ($J=1/2$) spin-orbital physics. Overall, the hierarchy of the determined multiplet parameters firmly locates the $4d^4$ ions in \KRC\ in the $LS$-coupling regime. The resulting $J=0$ singlet ground state and the weakness of interactions between the widely separated Ru ions explain the absence of structural (Jahn-Teller) and magnetic transitions in this compound.

The Hund's coupling $\JH=0.42$ eV for \KRC\ is larger than 0.34 eV for \CRO\ \cite{Gretarsson:prb:2019} and $\alpha$-RuCl$_3$ \cite{Suzuki:arxiv:2020}. This indicates the weaker screening effects in the large-gap Mott insulator \KRC. On the other hand, $\xi=0.103$ eV for \KRC\ is significantly smaller than 0.13 eV for \CRO\ and 0.15 eV for $\alpha$-RuCl$_3$. We ascribe this observation to the dynamical JT coupling. While the SOC renders the $J=0$ ground state JT inactive, the excitation energy of the $J=1$ triplet states is lowered by the JT stabilization energy \cite{Abragam::1970}. This leads to an apparent reduction of the $\xi$ parameter. This reduction, which acts in addition to the covalency effect, is special to \KRC, where the Ru ions weakly interact and are subject to single-ion JT dynamics in an ideal cubic environment.

Our result has broad consequences for the description of magnetism in $4d^4$ Ru-based Mott insulators. We have clearly revealed the $J$ multiplet formation in the cubic crystal field environment, highlighting the unquenched orbital momenta actuated by the intraionic SOC. Existing descriptions of the singlet ground states in ruthenium magnets \cite{Lee:natm:2006,Miura:jpsj:2007,Yun:prb:2019,Khalifah:sci:2002,Wu:prl:2006}, however, tend to \textit{a priori} assume spin-only $S=1$ states with quenched orbital moment due to crystal field distortions, and generally neglect the role of SOC. Our result, on the contrary, suggests that the $J$ multiplets are the generic starting point in the study of magnetism in Ru compounds and encourages reexamination of the theoretical description of the quantum singlet states in these materials.

We conclude by pointing out future directions of excitonic magnetism research in ruthenium compounds. The identification of the nonmagnetic end member \KRC\ and the ordered antiferromagnet \CRO\ naturally raises the interest in the quantum critical point separating the two phases, where the intensity of the amplitude fluctuations (Higgs mode) is most enhanced \cite{Jain:natp:2017}. Continuous tuning of the tetragonal distortion by uniaxial stress \cite{Hicks:sci:2014} could provide insight into the quantum critical behavior in the course of triplon condensation. Furthermore, the arrangement of $J$ multiplets on the honeycomb lattice induces highly frustrating, bond-dependent exchange interactions, which lead to exotic phenomena such as the nontrivial triplon topology \cite{Anisimov:prl:2019} or the bosonic analog of the Kitaev honeycomb model \cite{Chaloupka:prb:2019}. The honeycomb-lattice ruthenate Ag\tds{3}LiRu\tds{2}O\tds{6} is a candidate material realizing these phenomena \cite{Kimber:jmc:2010,Kumar:prb:2019}, which indeed does not exhibit magnetic order down to low temperature. Future investigations into ruthenium magnets on different lattices will further enrich the physics of excitonic magnetism.

In summary, we have demonstrated the formation of archetypal spin-orbit $J$ multiplets in the cubic \KRC\ by Ru $L_3$-edge RIXS. The observed multiplets  are well described by an ionic Hamiltonian without crystal field distortion, locating the 4$d^{4}$ ions in \KRC\ in the $LS$-coupling regime, with a $J=0$ singlet ground state and weakly interacting $J=1$ triplet excitations. These observations firmly establish the $J$ multiplet formation in the Ru-based insulators, substantiating the excitonic magnetism driven by the $J = 1$ triplon condensation in \CRO. The unquenched orbital contribution to the magnetism calls for reconsideration of the interpretations of quantum singlet formations in a variety of compounds based on Ru$^{4+}$ ions. This result will encourage further investigations into the quantum criticality, the nontrivial topology of the triplon bands, and excitonic magnetism in various crystal environments.

\textit{Acknowledgements.}---
We thank K. S. Rabinovich, A. V. Boris, K. Kitagawa, and R. Oka for useful discussions.
The project was supported by the European Research Council under Advanced Grant No. 669550
(Com4Com). We acknowledge DESY, a member of the Helmholtz Association HGF, for the provision of experimental facilities. The RIXS experiments were carried out at the beamline P01 of PETRA III at DESY.
H.T. acknowledges financial support from the Max Planck-UBC-UTokyo center for Quantum Materials. H.S. acknowledges financial support from the Alexander von Humboldt Foundation.

\end{document}